\documentclass[aps,prb,twocolumn,groupedaddress]{revtex4-1}
\usepackage[english]{babel}
\usepackage{graphicx}
\usepackage{color}
\usepackage[utf8]{inputenc}
\usepackage{pstricks,pst-grad,color}
\usepackage{graphicx,pifont,amssymb,german}
\usepackage{amsmath,amssymb}

\begin{document}

\title{Orbital Order, Metal Insulator Transition, and
  Magnetoresistance-Effect in the two-orbital Hubbard model} 

\author{Robert Peters}
\email[]{peters@scphys.kyoto-u.ac.jp}
\affiliation{Department of Physics, Kyoto University, Kyoto 606-8502, Japan}
\author{Norio Kawakami}

\affiliation{Department of Physics, Kyoto University, Kyoto 606-8502, Japan}
\author{Thomas Pruschke}
\affiliation{Department of Physics, University of G\"ottingen, 37077 Göttingen, Germany}

\date{\today}

\begin{abstract}
We study the effects of temperature and magnetic field on a
two-orbital Hubbard model within dynamical mean field theory.
We focus on the quarter filled system, which
is a special point in the phase diagram due to orbital degeneracy.
At this particular filling the model exhibits two different 
long-range order mechanisms, namely orbital order and
ferromagnetism. Both can cooperate but do not rely on each
other's presence, creating a rich phase diagram.
Particularly, in the vicinity of the phase transition to an orbitally
ordered ferromagnetic 
state, we observe a strong magnetoresistance effect. Besides the low
temperature phase transitions, we also observe
a crossover between a paramagnetic insulating and a paramagnetic
metallic state for increasing Hund's coupling at high temperatures.
\end{abstract}


\maketitle

\section{Introduction}
Strong correlations among the electrons and orbital degeneracy both
play a major role in the low temperature physics of transition
metal compounds \cite{maekawa2004}. Materials within this class
range from the recently discovered superconducting iron pnictides to
the manganites showing the colossal magnetoresistance.  

Due to the short spatial extension of the 3d-orbitals in the transition
metal atoms and the typical crystal structure of these compounds,
screening of the 
Coulomb interaction can be considered
weak, leading to the above mentioned strong local electron-electron
interactions.  
Within the group of transition metal
compounds, the cubic perovskite structure is a particularly common one.
In this structure the 5-fold degenerate d-orbitals split
into three-fold degenerate $t_{2g}$-orbitals and two-fold degenerate
$e_g$-orbitals. Therefore, besides strong correlations also orbital
degeneracy will play an important role in these
compounds. Especially near integer filling, orbital 
degeneracy can induce long-range orbital order
\cite{roth1966,kugel1973,tokura2000,fazekas2000,oles2010}, 
for which the expectation value to find an electron in one orbital depends
on the lattice site and the orbital. It can be
accompanied by a lattice 
distortion, Jahn-Teller distortion, caused by the coupling between
electrons and the lattice \cite{jahn1937,kanamori1960,gehring1975}.
However, even without such a lattice distortion it can be energetically
favorable to form an orbitally ordered state \cite{Leonov2008}.

Although the perovskite crystal structure is comparatively simple, a
full investigation of  
material-specific properties is still a big challenge. On the other
hand, as a rather large 
number of different compounds do show qualitatively similar physics,
it is suggestive 
to understand these common aspects by studying a model concentrating
on the most important 
ingredients of the 3d transition metal perovskites. As has been
discussed by a variety of authors, 
such a model is the multi-orbital Hubbard Hamiltonian
\cite{oles1983,imada1998}, which will therefore build the basis of our
investigations. 
In this article we especially want to focus on the interplay between
orbital degeneracy and 
strong correlations leading to either a competition or cooperation
between long-range ordered phases of spin, orbital and charge degrees
of freedom.  Particularly interesting for manganites is the case of
a twofold degenerate 
d-band at the Fermi energy, and here the special point of quarter
filling seems to play a major  
role for the physics of this class of compounds. Therefore, our aim
is to study the physics of the two-orbital Hubbard Hamiltonian at
quarter filling, with special emphasis laid on the phase diagram  
for the magnetic and the orbital order, and the changes in various
physical quantities across the phase boundaries. 

This article is organized as follows. After this introduction we will
specify the model, shortly explain the used methods, and
give a short overview about the ground state properties at
quarter filling. Thereafter we will study the influence of
temperature and of magnetic fields on the ordered phases.
As a particularly interesting quantity with respect to experiment we
will also present results for  the conductivity and its changes across
the phase boundaries. A summary will conclude the paper. 

\section{Model}
As already noted in the introduction, a reasonable qualitative
description of the low-energy properties of 3d-transition metal
perovskites can be obtained by the multi-orbital Hubbard model 
\cite{hubbard1963,kanamori1963,gutzwiller1963,oles1983} 
\begin{eqnarray}
H&=&H_T+H_U\nonumber\\
H_T&=&\sum_{<i,j>,\sigma,m}t_{i,j}c^\dagger_{i,\sigma,m}c_{j,\sigma,m}\nonumber\\
H_U&=&\sum_i\Bigl(\sum_m Un_{i,\uparrow,m}n_{i,\downarrow,m} \nonumber\\
&&+\left(U^\prime-\frac{J}{2}\right)\left(n_{i,\uparrow,0}+n_{i,\downarrow,0}\right)\left(n_{i,\uparrow,1}+n_{i,\downarrow,1}\right)
\nonumber\\
&&-2J\vec{S}_{i,0}\cdot\vec{S}_{i,1}\Bigr)\nonumber,
\end{eqnarray}
where $c^\dagger_{i,\sigma,m}$ creates an electron at site $i$, with
spin $\sigma$ in orbital $m$. Furthermore, $n_{i,\sigma,m}=c^\dagger_{i,\sigma,m} c_{i,\sigma,m}$ is
the density operator and $\vec S = c^\dagger_{\sigma_1} \vec
\sigma_{\sigma_1,\sigma_2} c_{\sigma_2}$ is the spin operator for the electrons.
$H_T$ corresponds to a hopping of the electrons between nearest neighbor
sites and $H_U$ represents a pure local two-particle interaction. 
The interaction consists of an intra-orbital density-density
interaction $U$, an inter-orbital density-density interaction
$U^\prime-J/2$, as well as a ferromagnetic Hund's coupling between the
orbitals, $-2J$.

We here neglect the pair-hopping term in the Hamiltonian, which should
have only minor quantitative and no qualitative influence as we perform
calculations for strong repulsive $U$ and away from
half filling. Nevertheless, it should be stated that there is no
rotational orbital symmetry due to the exclusion of the pair-hopping
term. However, due to this approximation, it is possible to include
the orbital occupation as conserved quantum number into the
calculations, which considerably simplifies and speeds up the
numerical calculations.  
To check the validity of this approximation, we have performed a few
additional calculations including the pair-hopping term, but no
significant differences have been found. 

Although including only local interactions and nearest neighbor
hopping terms, the Hubbard model is very challenging.
For calculating the magnetic and orbital phase diagrams, we use the dynamical mean field
theory (DMFT) \cite{metzner1989,pruschke1995,georges1996}. Capturing
the local physics correctly, it has proved to be a 
very powerful instrument for analyzing and understanding strong
correlation effects. Moreover, although for long-range order it is
closely connected to standard mean field theory, the inclusion of
local dynamical properties significantly renormalizes the physical
properties, and even completely suppresses ordering where static mean
field approaches would predict some. Therefore, even if we cannot
account for spatial fluctuations properly, the DMFT
results will give a reasonable qualitative and thermodynamically
consistent account of possible phases and becomes exact in the
limit of infinite spatial dimensions. As we are in this
study interested in
the fundamental aspects of the interplay between orbital degeneracy
and strong correlations, DMFT is well suited. For a realistic
comparison to transition metal oxides, e.g. Manganites, the lattice
structure and also phonon modes should be taken into account.

The remaining obstacle is to solve the DMFT self-consistency equations, which are tantamount to
calculate spectral functions for a multi-orbital single impurity Anderson model \cite{anderson1961,hewson,pruschke1995,georges1996}. As we want to concentrate on $e_g$-type systems with a two-fold degenerate ground state multiplet, 
we can employ the Numerical Renormalization Group
(NRG) \cite{wilson1975,bulla2008}, which is able to reliably calculate
spectral functions at low and zero temperature also including spin-
and orbital-symmetry-broken states. To be able to calculate spectral
functions at arbitrary temperatures within NRG, we 
use the complete Fock space algorithm \cite{peters2006,weichselbaum2007}.

An important aspect of the DMFT is that the lattice structure enters
only through the non-interacting density of states (DOS). Furthermore,
apart from quantitative details, the basic physical properties are
rather insensitive to the actual form of the DOS, as long as
particle-hole symmetry holds. As we are interested
in qualitative aspects of the physics of the two-orbital Hubbard
model, we therefore have a certain freedom to choose a numerically
convenient DOS. 
We thus choose a semi-elliptic local density of states with
bandwidth $W=4t$ for the non-interacting system. 
For the NRG calculations we use
$N=4000$ to $N=5000$ states kept per NRG iteration and an NRG
discretization parameter $\Lambda=2.0$. \cite{bulla2008}
Throughout this article the local Hubbard interaction is set to
$U/W=4$, which is a reasonable value for transition metal oxides.

\section{Ground state phase diagram}
Ferromagnetism and orbital order in a
multi-orbital Hubbard model in infinite dimensions was analyzed by
other authors before 
\cite{held1998,momoi1998,sakai2006,sakai2007,chan2009,kita2009}.
However, up to now the connection between the itinerant ferromagnetic phase,
most pronounced for a filling $n=1.4 $, and the orbitally ordered
ferromagnetic phase at quarter filling was not sufficiently analyzed.

\begin{figure}[htp]
\begin{center}
\includegraphics[width=0.8\linewidth,clip]{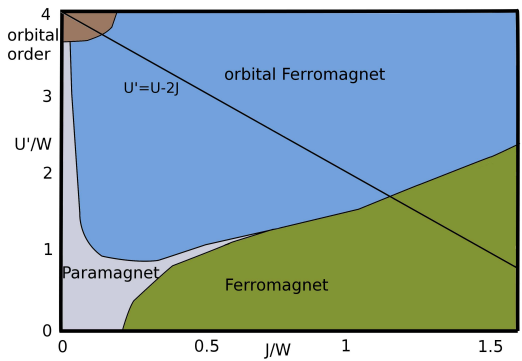}
\caption{(color online) Schematic ground state phase diagram at
  quarter filling for $U/W=4$. The black
  line represents $U^\prime=U-2J$.
\label{groundstate_quarter}} 
\end{center}
\end{figure}

The ground state phase diagram of the two-orbital Hubbard model was recently 
analyzed in some detail by \textcite{peters2010}. In particular, for
quarter filling the schematic phase diagram 
shown in Fig. \ref{groundstate_quarter} was obtained. 
In this figure both interaction parameters
$U^\prime$ and the Hund's coupling $J$ are treated as independent
parameters. The black 
line represents the relation $U^\prime=U-2J$ corresponding to 
the orbital symmetric case, in which the atomic ground state at
half filling is a spin triplet. 
The ground state phase diagram consists of 4 different phases. 
For the case of both interaction parameters ($U^\prime$ and $J$) being
small, there exists only  a paramagnetic metallic phase. 
For small Hund's coupling $J$ but large repulsive inter-orbital density-density
interaction $U^\prime$, there appears an antiferro-orbitally ordered phase
without spin polarization: Like for the N\'eel state of the antiferromagnet, 
the lattice exhibits a bipartite AB-structure, where electrons on the
A-sites of the lattice only occupy 
the orbital $m=1$, while on the B-sites of the lattice they only
occupy the orbital $m=2$.
If, on the other hand, $U'$ is kept small and the Hund's coupling $J$
is increased, one 
encounters for $J/W\gtrsim0.3$ a homogeneous ferromagnetic phase
without orbital polarization. 
This ferromagnetic phase extends  to a filling of approximately
$n\approx 1.5$.\cite{peters2010} 
The bulk part of the phase diagram  at quarter filling with $J>0$ and
$U^\prime-J/2 >0$ 
is, however, an anti-ferro orbitally ordered, ferromagnetic state. Thus, both types of order are quite obviously supporting each other.
We should note here that for attractive inter-orbital interaction
$U^\prime-J/2<0$ the 
energetically lowest state is a charge-density wave \cite{peters2010},
which was however excluded in the phase-diagram Fig.\ \ref{groundstate_quarter}.
While the orbitally ordered state at $T=0$ always
results in an insulating state, the homogeneous ferromagnetic state and
the paramagnetic state both are metallic. We thus encounter a metal
insulator transition (MIT) at $T=0$ when changing the
interaction parameters.   

It is well known that strong correlations tend to localize electrons
at integer fillings
\cite{han1998,koga2002,pruschke2005,koga2005,werner2009} and that the
transition point in the 
two-orbital Hubbard model depends on the strength of the Hund's
coupling $J$. \cite{medici2010} Furthermore, the localized electrons in the insulating
region tend to form long range order.
At quarter filling for the two-orbital Hubbard model on a bipartite
lattice, this can be most  
efficiently be done by forming an orbital ordered ferromagnetic state,
thus decreasing the energy of the system. 

Comparing to paramagnetic results, the paramagnetic MIT occurs
approximately at the same value of Hund's 
coupling as this ferromagnetic MIT. However, the temperature scale of
the paramagnetic MIT will be smaller, i.e. the
paramagnetic MIT will be covered by ferromagnetic states having the
lower energy for a bipartite lattice.

If this MIT will be observable in experiments
is at least doubtful for two reasons. Firstly, it is not clear how one can 
experimentally change the inter-orbital interactions $J$ and
$U^\prime$ without also significantly changing the Hubbard interaction
$U$. Secondly, the neglected coupling to the lattice, which
enhances the tendency towards orbital order due to
Jahn-Teller-distortions, will quite likely cover this purely electronic effect. Nevertheless, it is important to realize that
orbital order can be stabilized without resorting to lattice degrees
of freedom by correlation effects only.  
\section{Influence of temperature}
\begin{figure}[htp]
\begin{center}
\includegraphics[width=1\linewidth,clip]{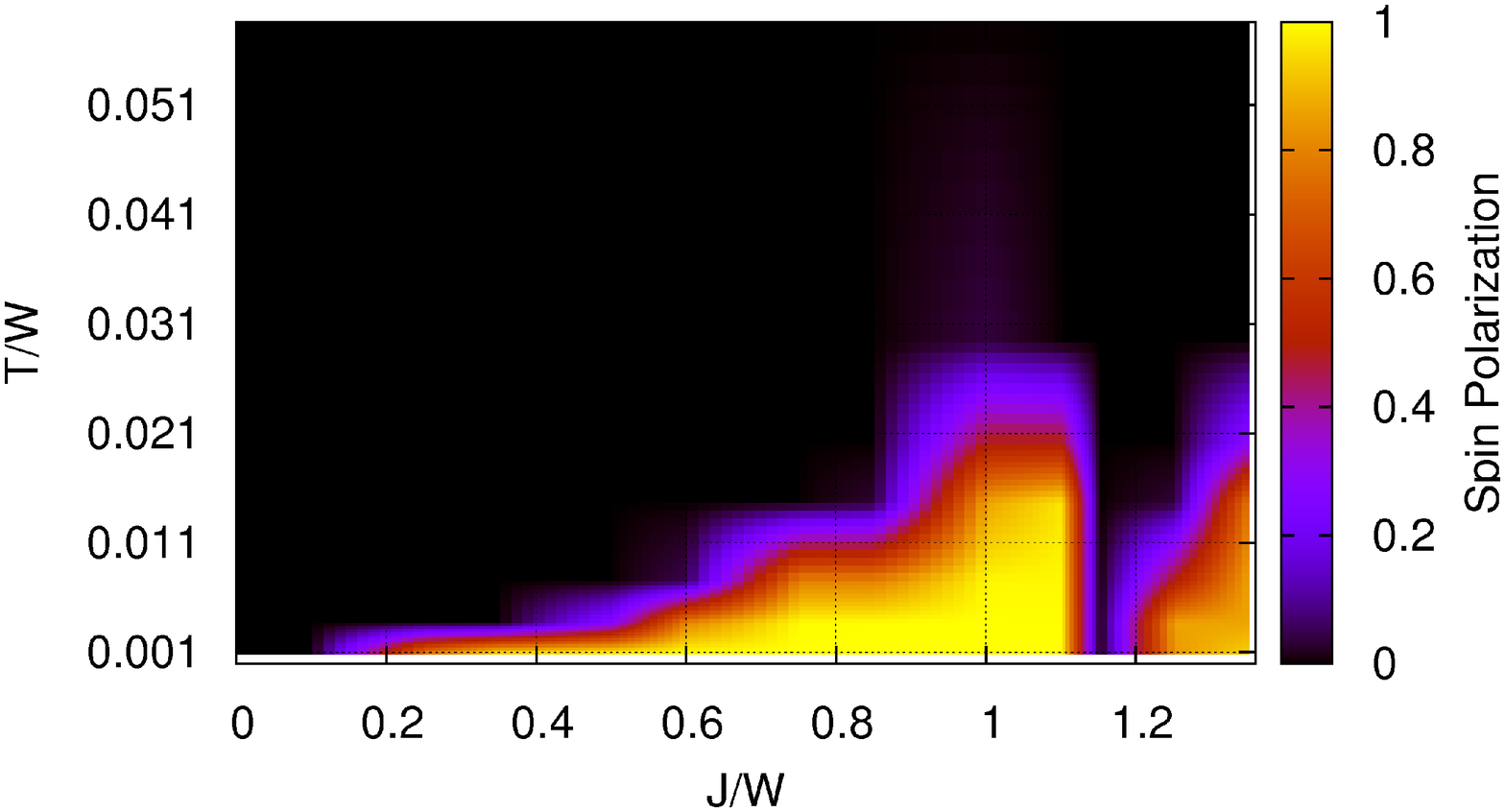}
\includegraphics[width=1\linewidth,clip]{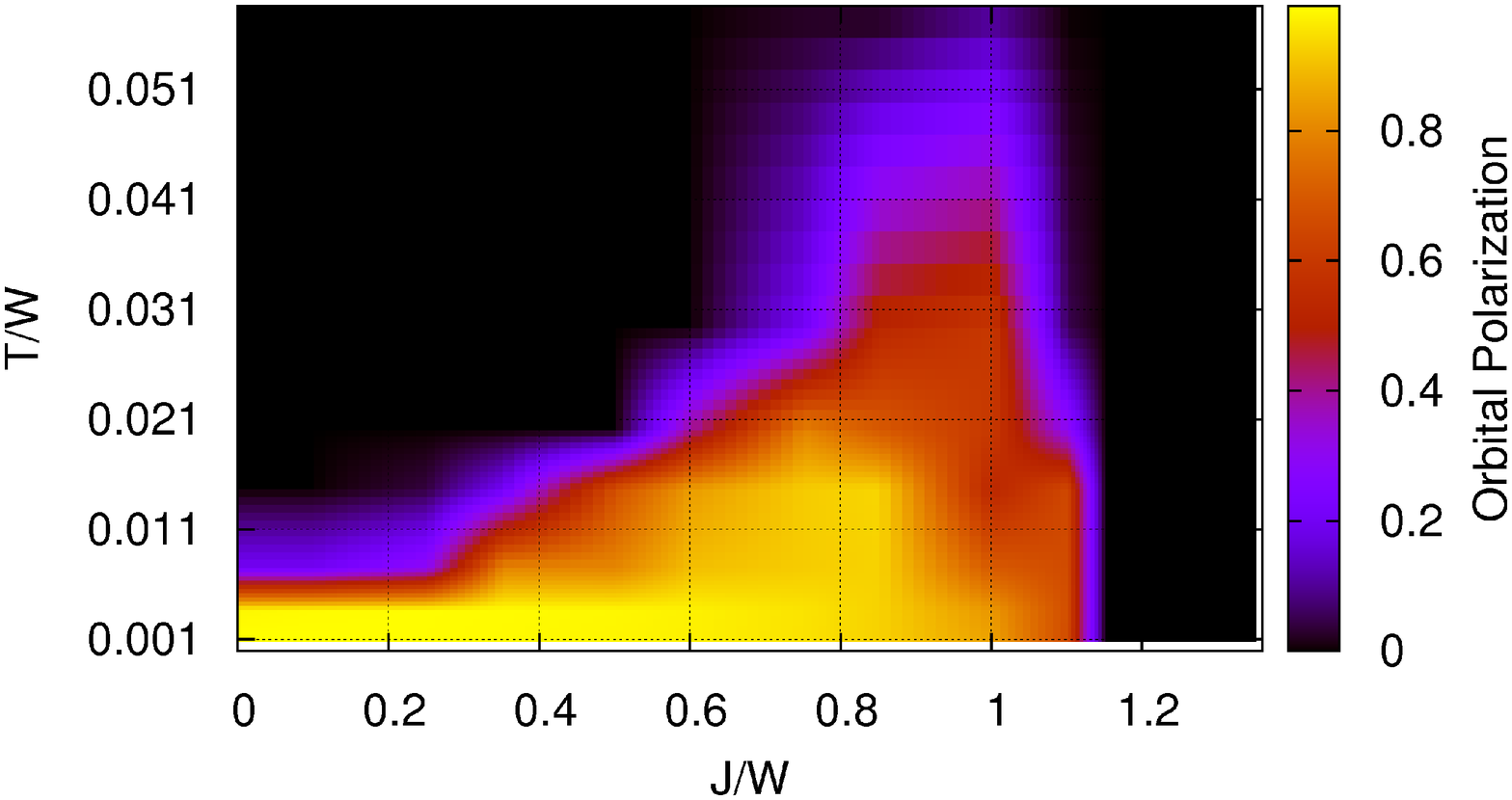}
\caption{(color online) Spin polarization ($M=\sum_m
(n_{m,\uparrow}-n_{m,\downarrow})$:  upper panel) and orbital
polarization ($P=\sum_\sigma (n_{1,\sigma}-n_{2,\sigma})$: lower
panel) for different temperatures and 
Hund's coupling at quarter filling and $U^\prime=U-2J$, $U=4W$. 
 The dip at $J/W\approx 1.2$ in the magnetic polarization
  corresponds to the transition between the orbitally ordered
  ferromagnet for $J/W<1.2$ and 
  the homogeneous ferromagnet for $J/W>1.2$. 
\label{magnetic_polar}}
\end{center}
\end{figure}
Experimentally more relevant than changing 
the interaction parameters
is the dependence of physical properties on temperature and magnetic field.
In order to limit the number of parameters in our system, we will from now on
focus on the special combination $U^\prime=U-2J$ for the interaction
parameters, representing the SO(3) symmetry for isolated atoms
\cite{oles1983}. Figure \ref{magnetic_polar} shows 
the results for the spin and orbital 
polarization as function of temperature $T$ and Hund's coupling $J$
in a false color plot. 
As we have only a limited number of actual data available, we used a
polynomial fit to obtain a reasonably smooth density plot. Thus, the
precise values and details of the transition lines should be regarded
with some caution.

However, the temperature gap in the spin
polarization at $J/W\approx 1.2$ is real and corresponds to the transition between orbitally
ordered ferromagnetism for $J/W<1.2$ and homogeneous ferromagnetism (without
orbital order) for $J/W>1.2$. In the former case, the ferromagnetic state is stabilized by
orbital order induced by strong $U^\prime$, which is an insulator and consequently leads to the formation of local moments, which then order due to double exchange mechanism \cite{zener1951,anderson1955,degennes1960}. For $J/W>1.2$ the orbital order has vanished (see lower panel of Fig.\ \ref{magnetic_polar}), but
one still finds a ferromagnetic solution, however with reduced Curie temperature and magnetic moment. The magnetic state in this parameter region remains metallic and must therefore be characterized as itinerant ferromagnet stabilized
mainly due to a gain in kinetic energy. 
These two rather different ferromagnetic phases are separated by a
small region with a paramagnetic and orbitally disordered metal. 

\begin{figure}[htp]
\begin{center}
\includegraphics[width=0.8\linewidth,clip]{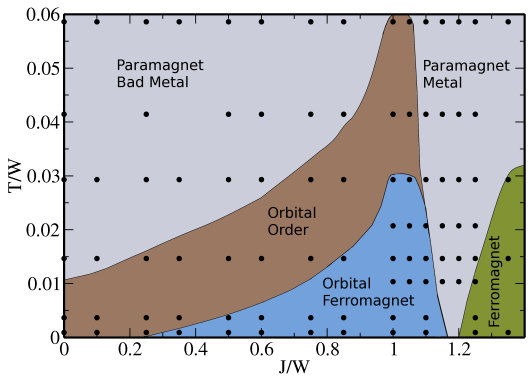}
\caption{
(color online) Schematic phase diagram for different temperatures and
  Hund's coupling at quarter filling and $U^\prime=U-2J$, $U=4W$. The
  black points denote the parameters at which calculations have been
  performed and from which the phase boundaries were drawn.
\label{TJ_phase}}
\end{center}
\end{figure}

From Fig. \ref{magnetic_polar} we have deduced the  schematic
 phase diagram presented in  Fig.\ \ref{TJ_phase}. The black points in the graph
 correspond to the parameters, at which actual calculations have been performed.
The orbital order is mainly driven by the inter-orbital density-density
interaction $U^\prime-J/2$. For small Hund's coupling, meaning strong
$U^\prime$, the transition temperature should behave like $T_c\sim
1/U^\prime$ in leading order from perturbation theory due to the same
arguments as for the antiferromagnetic N\'eel state at half filling for
large $U$.  Increasing Hund's coupling leads to a decreasing
$U^\prime$ and vanishing orbital order.
For $J>0$ and low temperatures the orbital order comes along with
ferromagnetic order.
\begin{figure}[htp]
\begin{center}
\includegraphics[width=0.8\linewidth,clip]{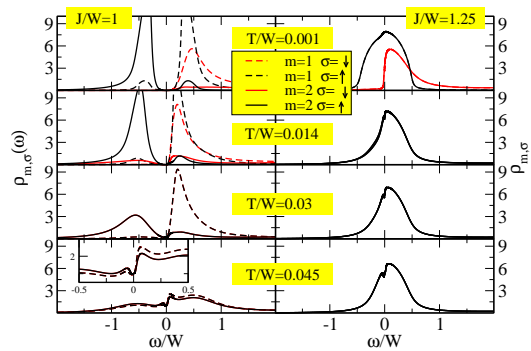}
\caption{(color online)
Left panels: Spectral functions within the orbitally ordered ferromagnetic region
for $U=4W$, $J/W=1$. For decreasing temperatures a gap at the
Fermi-energy $\omega=0$ is formed, and the spectral functions for $m=1$
and $m=2$ differ from each other. For temperatures below $T/W<0.02$
the orbitally ordered ferromagnetic state is stabilized. The
  inset in lowest panel shows a magnification around the Fermi
  energy clarifying the orbital order for this temperature.
Right panels: Spectral functions for $J/W=1.25$. There is no
difference in the spectral functions between $m=1$ and $m=2$.
Additionally, the ferromagnetic transition temperature
is much lower.
\label{Spectral}}
\end{center}
\end{figure}

Up to now we have only discussed the static properties in the different phases, frequently referring to one case as an insulator, the other as a metal. To explain this distinction we discuss now the
spectral functions for generic parameter values in the different cases.
Figure \ref{Spectral} shows spectral functions corresponding to the
cases $J/W<1.2$ (left panels) and $J/W>1.2$ (right panels). 
As already observed for the order parameters, one first finds orbital order with decreasing temperature for $J/W<1.2$, and a second transition 
to a ferromagnet for even lower temperatures.
In the case $J/W>1.2$, on the other hand, there is no sign for orbital
ordering, but a clear ferromagnetic polarization of the spectra for
the lowest shown temperature.
The spectral functions of both systems differ very strongly. The
orbitally ordered system exhibits  characteristic van Hove
singularities due to the reduced translational symmetry, and a
gap at the Fermi energy $\omega=0$, thus being an insulator. The system without
orbital order does not show this gap, but a smooth and finite DOS at
the Fermi energy, thus being metallic.

\begin{figure}[htp]
\begin{center}
\includegraphics[width=0.8\linewidth,clip]{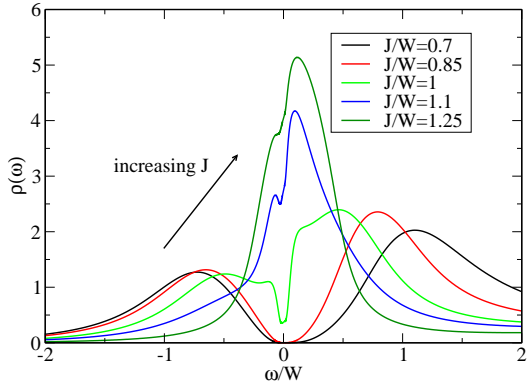}
\caption{
(color online) Spectral functions for $U=4W$, $T/W=0.08$,
  $U^\prime=U-2J$, and different $J$. For all shown spectral functions
  the system is in a paramagnetic state. For increasing $J$, the gap at
  the Fermi-energy $\omega=0$ vanishes.
\label{SpecPMIT}}
\end{center}
\end{figure}

However, the insulating behavior for $J/W<1.2$ is not only due to
orbital order, resulting in a doubling of the unit cell.
The gap can also be found 
above the critical temperature of the phase transition, see Fig.
\ref{SpecPMIT}. This figure shows spectral functions for $T/W=0.08$ in
the paramagnetic phase for different values of the Hund's coupling. By
increasing the Hund's coupling the gap is closed and a broad peak is
formed. At this temperature, there is no sharp phase transition
between the insulator and the metal, but a crossover
between both phases.  
One can argue that the gap is formed by the strong
inter-orbital density-density interaction. This emphasizes the
importance of taking into account the orbital level structure of Transition
Metal Oxides when modeling their properties, like the well-known
metal insulator transition in V$_2$O$_3$ \cite{imada1998}.

\section{Magnetoresistance}
\begin{figure}[htp]
\begin{center}
\includegraphics[width=1\linewidth,clip]{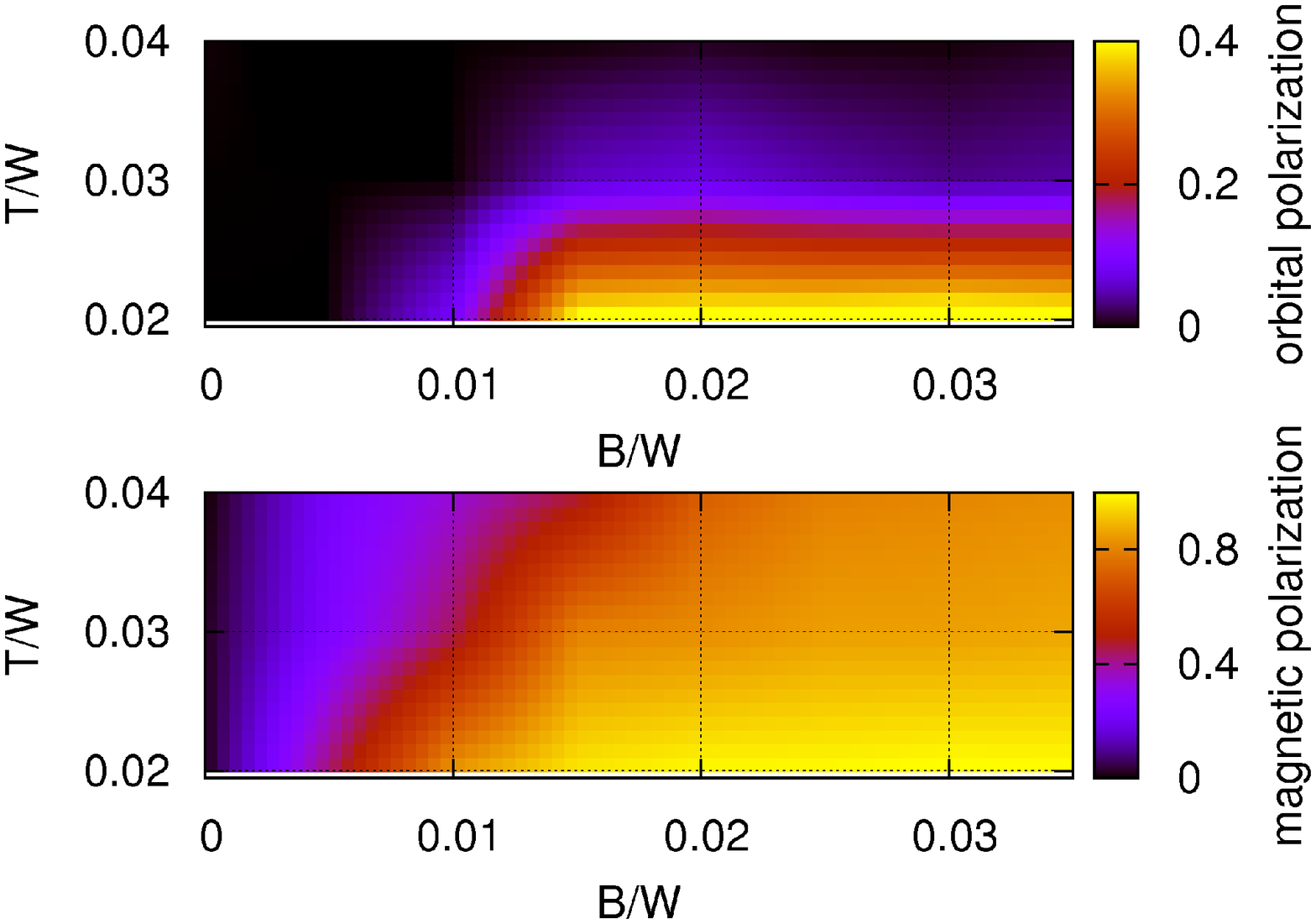}
\includegraphics[width=0.8\linewidth,clip]{conduct034.eps}
\caption{(color online) Response of the system to an applied magnetic
  field at $U/W=4$, $J/W=1.2$, $U^\prime=U-2J$.
The upper two panels show the magnetic and orbital polarization for
different temperatures $T/W$ and magnetic fields $B/W$.
The lower panels show the conductivity $\sigma$ scaled by
$\sigma_{max}=110$ (compare $\sigma_{max}$ with
Fig. \ref{magnetic_polarB038}) for $4$ different temperatures (lines
are meant as guide to the eye), and the
spectral functions at $T/W=0.02$.
\label{magnetic_polarB034}}
\end{center}
\end{figure}
\begin{figure}[htp]
\begin{center}
\includegraphics[width=1\linewidth,clip]{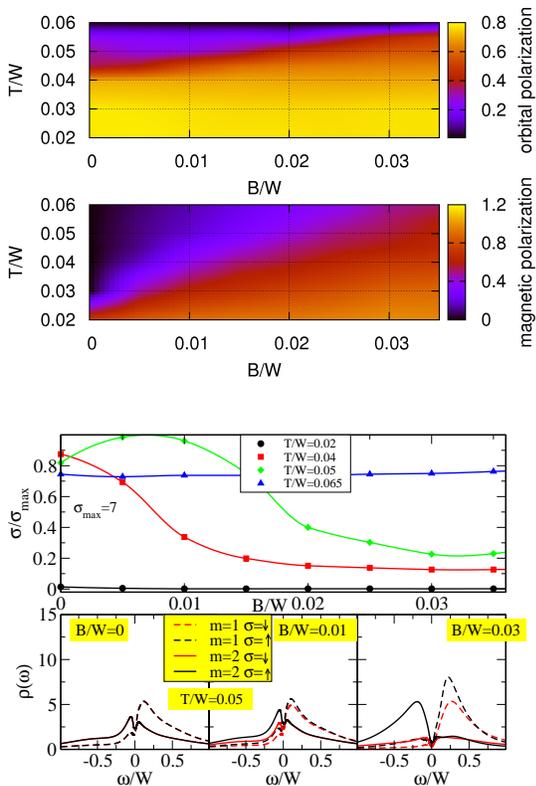}
\includegraphics[width=0.8\linewidth,clip]{conduct038.eps}
\caption{
(color online) Response of the system to an applied magnetic
  field at $U/W=4$, $J/W=1.05$, $U^\prime=U-2J$.
The upper two panels show the magnetic and orbital polarization for
different temperatures $T/W$ and magnetic fields $B/W$.
The lower panels show the conductivity $\sigma$ scaled by
$\sigma_{max}=7$ for $4$ different temperatures (lines
are meant as guide to the eye), and the
spectral functions at $T/W=0.05$.
\label{magnetic_polarB038}}
\end{center}
\end{figure}
Besides temperature, also the magnetic field can be varied very easily
in experiments. An increasing magnetic field will enhance
the magnetic polarization of the system. However, especially 
in the vicinity of the phase
transition $J/W\approx 1.2$, where the orbitally ordered ferromagnetic
state has disappeared, it is a priori not clear how the ferromagnetic
state 
will look like. When applying a magnetic field within the paramagnetic
phase between both ferromagnetic phases, the system can form either an
orbitally ordered or an orbitally homogeneous state. Both states do
considerably differ in their physical properties, as the orbitally
ordered state is gapped at the Fermi energy, while the homogeneous state
is not. 

In Figs. \ref{magnetic_polarB034} and \ref{magnetic_polarB038} we
analyze the conductivity of the system for different temperatures and
magnetic fields. The conductivity can be easily calculated within the
DMFT, as the self-energy is purely local. We here must take care of a
possible AB-ordering (N\'eel order) in the orbital index. The
conductivity at $\omega=0$ can be written as \cite{pruschke2003}
\begin{eqnarray}
\sigma(\omega=0)&=&c\sum_{\sigma,m}\int_{-\infty}^\infty d\epsilon \langle v^2\rangle
(\epsilon)\int_{-\infty}^\infty d\omega^\prime \frac{d
  f(\omega^\prime)}{d\omega^\prime}\nonumber\\
&&\cdot\bigl[A^{\mathcal{A}}_{m,\sigma}(\epsilon,\omega^\prime)A^{\mathcal{B}}_{m,\sigma}(\epsilon,\omega^\prime)+B_{m,\sigma}(\epsilon,\omega^\prime)^2\Bigr]\nonumber\\
A^{\mathcal{S}}_{m,\sigma}(\epsilon,\omega^\prime)&=&-\frac{1}{\pi}\mathfrak{Im}\Bigl(\frac{\zeta^{\mathcal{\bar
      S}}_{m,\sigma}}{\zeta^{\mathcal{A}}_{m,\sigma
}
  \zeta^{\mathcal{B}}_{m,\sigma}-\epsilon^2}\Bigr)\nonumber\\
B_{m,\sigma}(\epsilon,\omega^\prime)&=&-\frac{1}{\pi}\mathfrak{Im}\Bigl(\frac{\epsilon}{\zeta^{\mathcal{A}}_{m,\sigma}
  \zeta^{\mathcal{B}}_{m,\sigma}-\epsilon^2}\Bigr)\nonumber\\
\zeta^{\mathcal{S}}_{m,\sigma}&=&\omega^\prime+\mu-\Sigma_{m,\sigma}^{\mathcal{S}}(\omega^\prime)\nonumber,
\end{eqnarray}
where $A_{m,\sigma}(\epsilon,\omega^\prime)$ and
$B_{m,\sigma}(\epsilon,\omega^\prime)$ are the diagonal and off-diagonal Green's
functions of the AB-lattice respectively, $f(\omega^\prime)$ is the
Fermi-function, $\mu$ the chemical potential,
$\Sigma^{\mathcal{S}}_{m,\sigma}(\omega^\prime)$ the local self-energy.
 $\mathcal{S}=(\mathcal{A},\mathcal{B})$ corresponds to the
sub-lattice, for which 
$\mathcal{\bar A}=\mathcal{B}$ holds. Finally, $\langle v^2\rangle (\epsilon)$
is the averaged squared Fermi-velocity, in which the lattice structure
enters \cite{bluemer2003}. 
The prefactor $c=\frac{e^2}{h}\frac{\pi^2}{a}$ consists of the
resistance quantum and a non-universal part, depending on the details of
the lattice. In the following, we use c=1 for convenience.

Figure \ref{magnetic_polarB034} summarizes the properties of the system for
$J/W=1.2$. The upper two panels of the figure show the magnetic and
orbital polarization, while the lower panel shows conductivity and
spectral functions for different temperatures $T$ and magnetic fields $B$.
These parameters correspond to the minimum of
the Curie-temperature (dip) in Fig. \ref{TJ_phase}, where
there is neither orbital nor
magnetic order for $B=0$. We discuss the conductivity instead of the resistivity here as we expect a MIT at some point, leading to a drop to zero in $\sigma$, which is easier to visualize than a divergence in $\rho$.

If one applies a magnetic field
at low temperatures, $T/W \approx 0.02$, both spin and
orbital polarizations are induced. Assuming a bandwidth of
  $W=1eV$, this corresponds to a temperature of $T\approx 240K$. The system can still gain more energy by
localizing the electrons on different orbitals in the presence of a
magnetic field. This magnetoresistance effect corresponds to a
  MIT triggered by applying a magnetic field.
Furthermore, these results agree with the fact that applying
  a magnetic field at low temperatures near a MIT in a one-band
  model favors the insulating solution \cite{laloux1994}.
Looking at the spectral functions for $B/W=0.01$, the
middle of the lower panels 
in the figure, a spin polarized state has
formed. Nevertheless, all spectral functions still have spectral weight at the
Fermi energy. The majority-spin spectral functions form a two peak
structure, firstly even increasing the weight at the Fermi
energy. Further increasing the magnetic field above $B/W>0.015$ (right part of lower panel)
stabilizes the usual orbitally ordered ferromagnetic state with a gap
at the Fermi energy. 

This behavior is reflected in the 
conductivity, which directly depends on the spectral weight at the
Fermi energy. With increasing magnetic field
the conductivity initially increases at low temperatures, but then drops to zero as the
gap opens. If one further decreases the temperature, the initial
conductivity increase for small magnetic fields will be more pronounced,
but it again eventually drops to zero for magnetic fields of approximately
$B/W\approx  0.015$. Thus, at low temperatures, one observes a rather
dramatic magnetoresistance effect as function of magnetic field,
reminiscent of the colossal magnetoresistance effect in the
manganites. 

For higher temperatures the situation is different. In
this case, no orbital polarization is stabilized, and no gap is formed
at the Fermi energy. Increasing the magnetic field causes the
conductivity to increase smoothly, as the spectral weight of the majority spin
at the Fermi energy is increased.

Let us compare these results with those for $J/W=1.05$ collected in Fig. \ref{magnetic_polarB038} .
For this value of $J$, the orbital order has its highest transition temperature. Nevertheless,
for temperatures of approximately the transition temperature, the
spectral function is non-zero at the Fermi energy, even in the orbitally
ordered state. To compare
the conductivity in Figs. \ref{magnetic_polarB034} and
\ref{magnetic_polarB038} properly, one should also compare the scaling
constants $\sigma_{max}$. 
Increasing $J$ leads to a more pronounced peak at the Fermi energy in
the spectral function. 
This is reflected in $\sigma_{max}$, which for
$J/W=1.2$ takes the value $\sigma_{max}=110$ (in arbitrary 
units, neglecting the constant factors $c$ for calculating the
conductivity, which is the same in both cases), while for the orbitally
ordered ferromagnetic ground state the scaling constant is
$\sigma_{max}=7$. 

Increasing the temperature without applying a
magnetic field, leads first to a vanishing spin polarization  and
later, at
temperatures $T/W\approx 0.05$, to a paramagnetic state. For
$T/W<0.04$ the system is gapped and the conductivity is nearly
zero. Approaching 
now the transition temperature of the orbitally ordered phase, spectral weight
is shifted to the Fermi energy. At this point, a magnetic field leads
again to an increase of the gap width and to a stabilization of the
orbital order. The corresponding temperature for the
  magnetoresistance-effect is $T\approx 600K$, again assuming a
  bandwidth $W=1eV$.

This behavior can be seen in the upper panels of Fig.
\ref{magnetic_polarB038} in which the transition temperature of the
orbital order increases with increasing magnetic field, and in the
lower panels directly for the spectral functions or for the conductivity. For $T/W\approx 0.05$, the latter now again shows a maximum at some small magnetic field, decreasing again with increasing magnetic field. However, compared to the case $J/W\approx1.2$, the effect here is less dramatic.
For even higher temperatures, neither an orbital polarization nor
a gap at the Fermi energy is induced by a magnetic field resulting in
a nearly constant conductivity. 

One can do similar calculations for smaller Hund's coupling
$J/W<1$ or larger $J/W>1.25$.  For smaller Hund's coupling the
system is in an insulating state even for temperatures above the critical temperature for the 
ordered phases. Applying magnetic fields will also increase the transition
temperature of the orbitally ordered phase, but as the system is already
gapped in the paramagnetic state, the conductivity will not change
significantly. 
For larger Hund's coupling, applying a magnetic field will not induce
orbital order anymore. The system will stay in a metallic state with a correspondingly smooth and weak variation of the physical properties.

\section{Conclusions}
We have analyzed the properties of the two-orbital Hubbard model at
quarter filling for a range of temperatures and magnetic fields. As
expected, ferromagnetic order is prevalent at quarter filling, but
depending on the ratio of inter-orbital interaction and Hund's
exchange, it can appear either without or with accompanying orbital
order. 

Our calculations show that orbital order can be stabilized without including Jahn-Teller distortions,\cite{Leonov2008} and it is stabilized by a strong inter-orbital density-density interaction. If one
decreases the strength of this interaction, the orbital order will eventually vanish. As a result a metal insulator transition can be found, which can also be triggered by applying a magnetic field close
to the transition. This gives rise to a very strong magnetoresistance effect, as one can drive the system 
from a metallic into an insulating state.  The strength of this
magnetoresistance effect depends sensitively on the Hund's coupling $J$. In
particular, we found a critical regime for $J$ between an orbitally
ordered ferromagnet and an orbitally homogeneous one, where a
particularly strong magnetoresistance effect exists.  Note that these
interaction values for this particularly 
interesting behavior, $U/W=4$ and $J/W=1.2$, are not completely
unrealistic for Transition Metal Oxides. It therefore might be
interesting to look for 
$e_g$ systems at quarter filling which are in this parameter regime if
they exhibit strong magnetoresistance.

Another rather amusing fact is that we found an insulating paramagnetic state for strong
inter-orbital density-density interaction reaching to temperatures far
above the long-range ordered phases. This shows the importance of
the orbital level structure when analyzing the metal insulator
transition in, for example, V$_2$O$_3$.


\begin{acknowledgments}
RP thanks the Japan Society for the Promotion of Science (JSPS)
together with the Alexander von Humboldt-Foundation
for a postdoctoral fellowship. 
This work was supported by KAKENHI (Nos. 21740232, No. 20104010), the
Grant-in-Aid for the Global COE Programs ``The Next Generation of
Physics, Spun from Universality and Emergence'' from MEXT of Japan, and
the Funding Program for World-Leading Innovative R$\&$D on Science and
Technology (FIRST Program). TP acknowledges support by the German Science Foundation (DFG) through SFB 602.
Part of the calculations were
performed at Norddeutsche Verbund f\"ur Hoch- und H\"ochstleistungsrechnen (HLRN).
\end{acknowledgments}

\providecommand{\noopsort}[1]{}\providecommand{\singleletter}[1]{#1}%
\end{document}